\documentclass[11pt]{article}
\usepackage[utf8]{inputenc}
\usepackage[margin=1in]{geometry}

\usepackage{microtype}
\usepackage{graphicx}
\usepackage{subfigure}
\usepackage{booktabs} 
\usepackage{natbib}
\setcitestyle{open={(},close={)}}

\usepackage{amsmath}
\usepackage{mathtools}
\usepackage{amsthm}
\usepackage{algorithmic}
\usepackage{graphicx}
\usepackage{textcomp}
\usepackage{xcolor}
\usepackage{amsfonts}     
\usepackage{epsfig}
\usepackage{algorithm}
\usepackage{wrapfig}
\usepackage{balance}
\usepackage{url}
\usepackage{mathtools}
\usepackage{natbib}
\usepackage{multirow}
\usepackage{soul,comment}

\usepackage[textsize=tiny]{todonotes}

\usepackage[colorlinks=true,citecolor=blue]{hyperref}

\usepackage{amsmath,amsthm,amsfonts,amssymb,mathdots,array,mathrsfs,bm,bbm,stmaryrd,graphicx,subfigure,xcolor}

\usepackage{breakcites}

\usepackage[T1]{fontenc}
\usepackage{enumerate}
\usepackage{inputenc}

\usepackage{graphicx} % more modern
\usepackage{subfigure}

\usepackage{booktabs,balance}
\usepackage{rotating}
\usepackage{boldline}
\usepackage{makecell}
\usepackage{multirow}
\usepackage{balance}

\usepackage{tikz}

\newtheorem{definition}{Definition}[section]

\newcommand{\bsmat}{\begin{bmatrix} }
\newcommand{\esmat}{\end{bmatrix} }

\usepackage{environ}
\NewEnviron{smallequation}{%
    \begin{equation}
    \scalebox{0.97}{$\BODY$}
    \end{equation}
    }
    \NewEnviron{smallalign}{%
    \begin{equation}
    \scalebox{0.97}{$\BODY$}
    \end{equation}
    }

\begin{document}

\title{\bf\Huge Integrity Authentication in Tree Models}

\author{\vspace{0.5in}\\\textbf{Weijie Zhao, Yingjie Lao, Ping Li} \\\\
Cognitive Computing Lab\\
Baidu Research\\
10900 NE 8th St. Bellevue, WA 98004, USA\\\\
  \texttt{\{zhaoweijie12, laoyingjie,  pingli98\}@gmail.com}
}

\date{\vspace{0.5in}}
\maketitle

\begin{abstract}\vspace{0.3in}

\noindent Tree models are very widely used in practice of machine learning and data mining. In this paper, we study the problem of model integrity authentication in tree models. In general, the task of model integrity authentication is the design \& implementation of mechanisms for checking/detecting whether the model deployed for the end-users has been tampered with or compromised, e.g., malicious modifications on the model. We propose an authentication framework that enables the model builders/distributors to embed a signature to the tree model and authenticate the existence of the signature by only making a small number of black-box queries to the model. To the best of our knowledge, this is the first study of signature embedding on tree models. Our proposed method simply locates a collection of leaves and modifies their prediction values, which does not require any training/testing data nor any re-training. The experiments on a large number of public classification datasets confirm that the proposed signature embedding process has a high success rate while only introducing a minimal prediction accuracy loss.
\end{abstract}

\newpage

\section{Introduction}

Along with the evolution of machine learning, the growing cost of model building in terms of computational power, data annotation, and human expertise calls for intellectual property (IP) protection methods to secure the innovations and creative endeavors in this field. The problem is clearly aggravated by recent studies demonstrating that machine learning models are vulnerable to various categories of adversarial attacks~\citep{szegedy2014intriguing,papernot2016limitations,kurakin2016adversarial,liu2018survey,yuan2019adversarial,doan2021back,doan2021lira} and model stealing via reverse engineering or model extraction~\citep{hua2018reverse,hu2019neural,juuti2019prada}. Especially under the machine learning as a service (MLaaS) paradigm where the supply chain of models may involve multiple parties and vendors, the proprietary assets of the model IP owner, including data, algorithm, and computation infrastructure, are vulnerable to breach. On the other hand, maliciously altered models, e.g., by poisoning or backdoor attacks~\citep{chen2017targeted,munoz2017towards,steinhardt2017certified,shafahi2018poison,liu2017neural,liu2017trojaning,gu2019badnets,doan2021back,doan2021lira}, will also impair the integrity, reputation, and profit of the model owner. These attacks have been extensively studied on various machine learning models, including collaborative filtering~\citep{li2016data}, logistic regression~\citep{munoz2019poisoning}, clustering~\citep{biggio2018data}, support vector machine~\citep{biggio2012poisoning}, and also deep neural networks~\citep{DBLP:journals/corr/YangWLC17,gu2017badnets,turner2019label,DBLP:conf/aaai/SahaSP20,shafahi2018poison}. Very recently, such attacks on tree models have also been investigated~\citep{chen2019robust,andriushchenko2019provably,grari2020achieving,zhang2020efficient}, which further demands techniques for protecting tree models.

\vspace{0.1in}

To this end, signature embedding and fingerprinting have emerged as promising directions for protecting the advanced machine learning models~\citep{uchida2017embedding,adi2018turning,zhang2018protecting,li2019prove,li2019persistent,rouhani2018deepsigns,fan2019rethinking,cao2021ipguard,le2020adversarial,zhong2020protecting,yang2021robust,lao2022identification,lao2022deepauth}. Conceptually, these techniques embed a unique signature into a model that will behave differently from other models. The presence of such signatures should be easily verifiable in a later stage, which can serve as an ownership or integrity proof. However, almost all of these prior methods are only targeting deep neural networks, which is understandable as deep neural networks have shown superior performance across many fields while requiring large costs to build. 

\vspace{0.1in}

In this paper, as opposed to focusing on deep neural networks, we take the first step to investigate signature embedding for (ensemble) tree models~\citep{Book:Friedman_83}. We follow the direction of embedding fragile watermarks~\citep{lao2022deepauth} for the purpose of verifying the integrity of the model, which is different from robust watermarks~\citep{uchida2017embedding,adi2018turning,zhang2018protecting,li2019prove,li2019persistent,rouhani2018deepsigns,fan2019rethinking,yang2021robust,lao2022identification} that are used to trace the IP ownership. In particular, we focus on boosted trees in our experiments, although the ideas and concepts are also applicable to (e.g.,) bagging~\citep{breiman1996bagging}. 

\subsection{Boosted Tree Models}

In machine learning practice, boosting~\citep{Article:Schapire_ML90,Article:Freund_95,Article:Freund_JCSS97,Article:Bartlett_AS98,Article:Schapire_ML99,Article:FHT_AS00,Article:Friedman_AS01,Proc:ABC_ICML09,Proc:ABC_UAI10,li2022fast} is one of the most successful learning paradigms. Boosting is typically integrated with trees~\citep{Book:Friedman_83} to produce powerful tools for classification and regression; Readers are also referred to some interesting discussions in 2010~\url{https://hunch.net/?p=1467}  for comparing deep neural networks with boosted trees. 

\vspace{0.1in}

As summarized in a recent paper on trees~\citep{fan2020classification}, there are two recent implementation tricks that have made boosted tree models considerably more practical:
\begin{itemize}
\item \citet{Proc:McRank_NIPS07} developed an adaptive binning strategy to effectively transform any numerical features into (non-negative) integer values. This binning operation is conducted on a feature-wise fashion hence it is highly efficient.  It is adaptive because, for each feature, it only assigns bins in the regions where there are data points. Recall that tree model~\citep{Book:Friedman_83} (recursively) partitions the data points in a feature-wise fashion.  Therefore, this simple trick has significantly simplified the implementation of trees and  at the same time also improved the efficiency. In addition, it also makes it convenient for parallelizing tree models.  

\item \citet{Proc:ABC_UAI10} derived the explicit (and robust) formula for tree-split criterion using the second-order gain information (i.e., the so-called ``Robust LogitBoost''), which typically improves the accuracy, compared to the implementation based on the criterion of using only the first-order gain information~\citep{Article:Friedman_AS01}. This formula for computing tree-split gain  has resolved the numerical issue in the original LogitBoost~\citep{Article:FHT_AS00,Article:Friedman_AS01,Article:FHT_JMLR08}. This formula has been widely used in practice and is the standard implementation in all major tree platforms. 
\end{itemize}

Another major progress is the so-called ``adaptive base class boost'' (ABC-Boost)~\citep{Proc:ABC_ICML09,Proc:ABC_UAI10} for multi-class classification by re-writing the derivatives of the classical multi-class logistic regression loss function. ABC-Boost often improves the accuracy of multi-class classification tasks, in many cases substantially so. 

\vspace{0.1in}

In this paper, we  develop integrity authentication schemes for ``Robust LogitBoost''~\citep{Proc:ABC_UAI10} and the code base uses the implementation of ``Fast ABC-Boost''~\citep{li2022fast}.

%\newpage\clearpage

\subsection{Challenges and Goals} 

Signature embedding or watermarking has been widely studied in the deep learning community~\citep{uchida2017embedding,adi2018turning,zhang2018protecting,li2019prove,li2019persistent,rouhani2018deepsigns,cao2021ipguard,fan2019rethinking,le2020adversarial,zhong2020protecting,yang2021robust,lao2022deepauth,lao2022identification}. Many of these works embed a desired behavior into the learned function by using backdoor techniques~\citep{adi2018turning} or enforcing a specific representation in a latent space~\citep{rouhani2018deepsigns}. However, similar techniques are not readily available for tree models in order to facilitate effective signature embedding, e.g., backdoor on tree models entails little study in the literature. In addition, tree models have different architectures and applications from deep neural networks, which also demands careful customization of tree model signature embedding. 

\vspace{0.1in}

In general, it is not straightforward to directly apply the signature embedding process of deep learning to tree models. We summarize the key challenges as follows:
\begin{itemize}
\item Deep learning methods require gradients. However, tree models are not differentiable.
\item Many deep learning signature embedding algorithms require to retrain the network. In the context of boosted tree models, each training iteration constructs a new tree that attempts to correct the misclassifications of previous trees. Appending more trees not only increases the model size (to store additional trees) but also damages the inference performance (to evaluate additional trees).
\item Meanwhile, retraining tree models by only replacing a subset of existing trees is still an open research direction since each tree is generated on the classification results of the previous trees---eliminating previous trees invalidates the boosting dependency. As a result, it is impractical to inplace retrain the tree models.
\end{itemize}
Therefore, this paper designs a new signature embedding framework for non-differentiable and non-inplace-retrainable tree models. We target to have a signature embedding framework to ``sign'' a trained tree model. After distributing the model to the host services, we are able to remotely access the model in a black-box fashion: we can query the model with several secret inputs (we call them \textit{signature keys}) and verify its output predictions. We have three main objectives: (i) The signature embedding procedure should be able to generate a vast number of distinct signs for different model host services. (ii) The signature can verify the integrity of the model that has not been tampered with. Whenever the model host modified the model, the output predictions of signature keys should be changed, i.e., the signature should be fragile against malicious modifications.  (iii) The signature embedding algorithm should not degrade the model prediction accuracy.

%On the other hand, various defensive methods have also been investigated to mitigate these threats. For causative attacks, most of prior works focused on reactive countermeasures, such as data sanitization~\citep{cretu2008casting,rubinstein2009antidote} and attack detection~\citep{munoz2017towards,DBLP:journals/corr/YangWLC17}. Different from the prior works, the proposed fragile signature based authentication can be effectively utilized as a proactive defense for ensuring the integrity of the model that has not been tampered with, since a maliciously altered model would not have the intact signature. 

\subsection{Summary of Contributions}
We summarize our contributions as follows:
\begin{itemize}
\item We introduce a novel model authentication framework and signature embedding algorithm for tree models. As far as we know, this is the first study that investigates model authentication on tree models.
\item We propose a novel searching and selection algorithm to generate signature keys and manipulate tree models.
\item We empirically evaluate the proposed algorithm on various public datasets. The results confirm the effectiveness and high authentication rate of our methods.
\item Since this paper is the first work to investigate signature embedding on tree models, we design baseline attacks to assess the quality of our technique, which also provides qualitative and quantitative metrics and comparisons for future tree model protection methods.
\end{itemize}

\textbf{Organization.} \ In the rest of the paper, we introduce our authentication framework in Section~\ref{sec:auth} and  present the signature embedding process in Section~\ref{sec:signature}. Section~\ref{sec:exp} reports the experimental evaluation. We present remarks and conclude this paper in Section~\ref{sec:conclusion}.

\section{Authentication Framework}\label{sec:auth} 

In this section, we first present the threat model for the signature embedding process. Then we describe the workflow of the signature enrollment and authentication.

\subsection{Threat Model}
 
The signature embedding process can be conducted by the model builder or a trusted party. Without loss of generality, we assume a pre-trained model could be received from a model builder who builds the model architecture and corresponding parameters with the training dataset, and a held-out validation dataset for evaluating the performance. We then apply the proposed methodology to this tree model to embed a desired signature. Only the legitimate model owner will have knowledge of the signature and the corresponding signature keys. In the authentication process, the model owner can \textbf{verify the presence of the signature by using the signature keys via the prediction API}---the model owner \textbf{only needs access to the predicted class during the authentication}. Thus, memorizing the tree weight hashing value or the predicted probability of some inputs may not be applicable. %The signature may also be used as an ownership verification. In this scenario, if the returned prediction of an unknown model is the same or very close to that of the model owner, this suspect model is likely pirated from the legitimate model owner. %Subsequently, the model owner may take legal actions after collecting other shreds of evidence for this matter.

%To compromise such signature embedding techniques, an adversary might attempt to remove the signature, while retaining the original functionality of the target model (i.e., the adversary still wants to use the model so that the evaluation performance should not be drastically failed). 
For signature embedding on DNNs, transformation attacks such as model compression, model fine-tuning, and signature overwriting are often used to evaluate the performance of an embedded signature. Since this paper is the first work to investigate signature embedding on tree models, we design similar transformation attacks as a baseline countermeasure to assess the quality of our proposed technique, which also provides qualitative and quantitative metrics and comparisons for future tree model protection methods. Note that the target of this work is a fragile signature, which could verify if a tree model has been tampered with and serve as a proactive defense against malicious model modifications, similar to~\cite{lao2022deepauth}. In other words, the objective of this signature is integrity verification as opposed to IP tracing as in some of prior DNN watermarking and fingerprinting methods~\citep{rouhani2018deepsigns,adi2018turning,cao2021ipguard,he2019sensitive,yang2021robust,lao2022identification}. Thus, we expect the embedded signatures to disappear after transformation attacks. Please also note that ambiguity attack is not a concern for fragile signature, as similarly laid out in~\cite{lao2022deepauth}. 
%Another concern raised recently for DNN watermarking is the ambiguity attacks~\citep{fan2019rethinking,li2019persistent}, which aim to cast doubt on the ownership verification by forging additional watermarks for a DNN model. Intuitively, if an adversary can embed a second watermark on a watermarked model, there is a huge ambiguity about the model's IP ownership. It is important to note that ambiguity attacks are not a concern for our proposed scheme. As the authentication is performed by checking the signature of an unknown model based on the enrolled information stored on the trusted server, a forged signature is extremely unlikely to match the legitimate signature, which hence does not create any ambiguity. %We assume the attacker is able to fully access to the model but does not know the watermark or the corresponding key samples.

\subsection{Flow of Enrollment and Authentication}\label{ssec:auth-flow}

%\hl{also need rephrase, update the notations, if use signature, it might be better to use s instead of m in the notation}

We formally express our proposed scheme as two phases:
\begin{itemize}
    \item Signature embedding $R^{\textit{msg}} \leftarrow \mbox{\bf Embed}(R,\textit{msg})$ that generates a signed version of regression trees $R^{\textit{msg}}$, signature key $\textit{key}$ from the given original regression trees $R$ with a target signature message $\textit{msg}$. %$k$ can be selected from $m$, which typically represents the signature keys that are used for verifying the signature.
    \item Signature extraction $\textit{msg}^{\prime} \leftarrow \mbox{\bf Extract}(R^{\textit{msg}},\textit{key})$ that collects the signature message $\textit{msg}^{\prime}$ using $\textit{key}$ from the given regression trees $R^{\textit{msg}}$. The generated $\textit{msg}^{\prime}$ will then be used for authentication. %If $m^{\prime}$ does not match the desired message $m$, it indicates that the integrity of the model has been tampered with.  %is a strong ownership proof, given that the signature embedding process ensures a low false positive rate.
\end{itemize}

\begin{figure*}[htbp]
\centering
\includegraphics[width=6.5in]{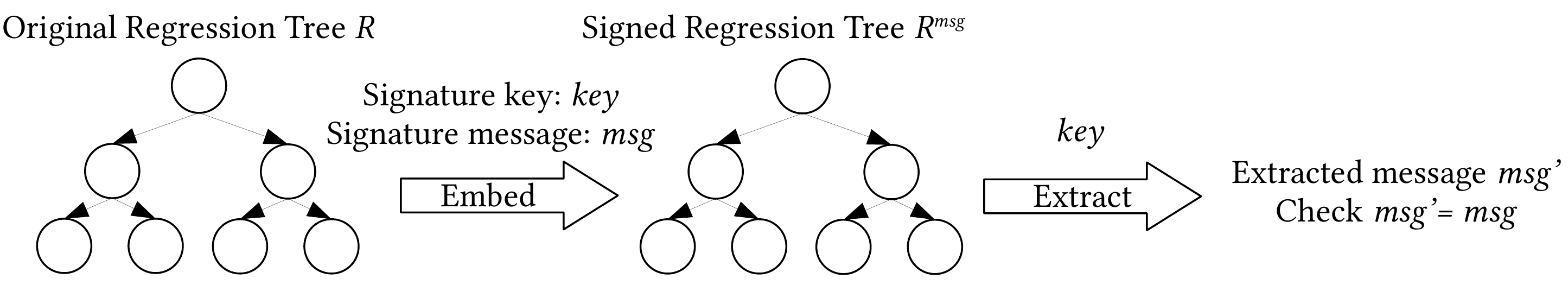}
\caption{Signature enrollment and authentication workflow.}\label{fig:workflow}
\end{figure*}

Figure~\ref{fig:workflow} presents a visual illustration of the signature enrollment and authentication workflow. 
%\hl{check notations and terminologies} 
Built upon the proposed signature embedding technique, an authentication process can be established. After successfully embedding the signatures, the signature key and the correct signature message are stored on a secure server. We can also choose to register the information of model ID during the enrollment to enable authentication for multiple legitimate users. In this case, each version of the model needs to be generated by applying the proposed technique with a different set of the signature key and signature message. Then, the model builder will sell, distribute, or deploy the model. Each customer may receive a different version of the original tree model (i.e., with a different signature). Later, when an authentication process is initiated, based on the model ID if it is enrolled, the server sends the signature key to the model API and collects the corresponding response. The server determines the authenticity of the unknown model by checking how this response matches the stored signature message. Thus, another advantage of this method is its small computational and communication overhead: only one authentication process, especially when comparing to those countermeasures that require periodically checking the accuracy~\citep{munoz2017towards,DBLP:journals/corr/YangWLC17}. We focus on the steps of embedding signatures on tree models in this paper. Modern authentication techniques and protocols~\citep{majzoobi2012slender,lee2014lightweight,alamr2018secure} can be implemented on top of the generated signatures.

\section{Signature Embedding}\label{sec:signature}

In this section, we begin with a brief introduction of boosted tree models as a preliminary. Then we present the proposed two-stage algorithm to embed signatures: 1)  locating the signature key candidates; 2) selecting independent signature keys and manipulating the prediction value on the terminal nodes.

\subsection{Boosted Tree Models: MART and Robust LogitBoost}

We denote a training dataset by $\{y_i,\mathbf{x}_i\}_{i=1}^N$, where $N$ is the number of training samples, $\mathbf{x}_i$ is the $i$-th feature vector, and  $y_i \in \{1, 2, ..., K\}$ is the $i$-th class label, where $K\geq 3$ in multi-class classification. We consider the same framework as in LogitBoost~\citep{Article:FHT_AS00} and MART (multiple additive regression trees)~\citep{Article:Friedman_AS01}, which can be viewed as generalizations to logistic regression and assume the class probabilities $p_{i,k}$  to be: 
\begin{align}\label{eqn_logit}
p_{i,k} = \mathbf{Pr}\left(y_i = k|\mathbf{x}_i\right) = \frac{e^{F_{i,k}(\mathbf{x_i})}}{\sum_{s=1}^{K} e^{F_{i,s}(\mathbf{x_i})}},\hspace{0.2in} i = 1, 2, ..., N,
\end{align}
where $F_{i,k}(\mathbf{x}_i)$ is an  additive model of $M$ terms: 
\begin{align}\label{eqn_F_M}
F^{(M)}(\mathbf{x}) = \sum_{m=1}^M \rho_m h(\mathbf{x};\mathbf{a}_m),
\end{align}
where  $h(\mathbf{x};\mathbf{a}_m)$ is typically a regression tree, and $\rho_m$ and $\mathbf{a}_m$ are parameters learned from the data, by minimizing the {\em negative log-likelihood loss}: 
\begin{align}\label{eqn_loss}
L = \sum_{i=1}^N L_i, \hspace{0.4in} L_i = - \sum_{k=1}^{K}r_{i,k}  \log p_{i,k}
\end{align}
where $r_{i,k} = 1$ if $y_i = k$ and $r_{i,k} = 0$ otherwise. The optimization procedures would need the derivatives of $L$ with respect to the function values $F_{i,k}$. The classical  multi-class logistic regression textbooks give
\begin{align}\label{eqn:logit_d1d2}
&\frac{\partial L_i}{\partial F_{i,k}} = - \left(r_{i,k} - p_{i,k}\right),
\hspace{0.5in}
\frac{\partial^2 L_i}{\partial F_{i,k}^2} = p_{i,k}\left(1-p_{i,k}\right).
\end{align}

\newpage

\begin{algorithm}[t]
\begin{algorithmic}[1]
\STATE $F_{i,k} = 0$, $p_{i,k} = \frac{1}{K}$, $k = 1$ to  $K$, $i = 1$ to $N$
\FOR{$m=1$ to $M$}
\FOR{$k=1$ to $K$}
  \STATE  $\left\{R_{j,k,m}\right\}_{j=1}^J = J$-terminal node regression tree from
 $\{r_{i,k} - p_{i,k}, \ \ \mathbf{x}_{i}\}_{i=1}^N$,  with weights $p_{i,k}(1-p_{i,k})$, using the tree split gain formula Eq.~\eqref{eqn:logit_gain}.
  \STATE $\beta_{j,k,m} = \frac{K-1}{K}\frac{ \sum_{\mathbf{x}_i \in
  R_{j,k,m}} r_{i,k} - p_{i,k}}{ \sum_{\mathbf{x}_i\in
  R_{j,k,m}}\left(1-p_{i,k}\right)p_{i,k} }$ 
\STATE $f_{i,k} = \sum_{j=1}^J\beta_{j,k,m}1_{\mathbf{x}_i\in R_{j,k,m}}$,\hspace{0.3in} $F_{i,k} = F_{i,k} + \nu f_{i,k}$
\ENDFOR
\STATE $p_{i,k} = \exp(F_{i,k})/\sum_{s=1}^{K}\exp(F_{i,s})$\\
\ENDFOR
\end{algorithmic}
\caption{Robust LogitBoost. MART is similar, with the only difference in Line 4.  }
\label{alg:robust_LogitBoost}
\end{algorithm}

Algorithm~\ref{alg:robust_LogitBoost} describes the details for {\bf Robust LogitBoost}~\citep{Proc:ABC_UAI10}. It fixes the previously thought ``numerical instability'' problem in the original LogitBoost paper~\citep{Article:FHT_AS00}, as discussed in~\cite{Article:Friedman_AS01,Article:FHT_JMLR08}. Robust LogitBoost~\citep{Proc:ABC_UAI10} uses the 2nd-order formula in~\eqref{eqn:logit_gain} for computing the gains when deciding the split  points.

Given $N$ data points which are assumed to be sorted according to the corresponding feature values. The tree-splitting procedure is to find the index $s$, $1\leq s<N$, such that the weighted  square error (SE) is reduced the most if split at $s$.  That is, we seek the $s$ to maximize
\begin{align}\label{eqn:logit_gain}
Gain(s) =&  \frac{\left[\sum_{i=1}^s \left(r_{i,k} - p_{i,k}\right) \right]^2}{\sum_{i=1}^s p_{i,k}(1-p_{i,k})}+\frac{\left[\sum_{i=s+1}^N \left(r_{i,k}- p_{i,k}\right) \right]^2}{\sum_{i=s+1}^{N} p_{i,k}(1-p_{i,k})}- \frac{\left[\sum_{i=1}^N \left(r_{i,k} - p_{i,k}\right) \right]^2}{\sum_{i=1}^N p_{i,k}(1-p_{i,k})}.
\end{align}
Because the computations involve $\sum p_{i,k}(1-p_{i,k})$ as a group, this procedure is actually numerically stable. In the implementation, to avoid occasional numerical issues, some ``damping'' could be added to the denominator, i.e., $\left\{\epsilon+\sum_{node} p_{i,k}(1-p_{i,k})\right\}$.

\vspace{0.1in}

In comparison, MART~\citep{Article:Friedman_AS01} only used the first order information to construct the trees, i.e.,
\begin{align}\label{eqn:mart_gain}
MartGain(t) =&  \frac{1}{s}\left[\sum_{i=1}^s \left(r_{i,k} - p_{i,k}\right) \right]^2+
\frac{1}{N-s}\left[\sum_{i=s+1}^N \left(r_{i,k} - p_{i,k}\right) \right]^2-\frac{1}{N}
\left[\sum_{i=1}^N \left(r_{i,k} - p_{i,k}\right) \right]^2.
\end{align}

\vspace{0.1in}

To avoid repetition, we do not provide the pseudo code for MART~\citep{Article:Friedman_AS01}, which is in fact almost identical to Algorithm~\ref{alg:robust_LogitBoost}. The only difference is in Line 4, which for MART becomes \\

$~\ $\hspace{0.5in}$\left\{R_{j,k,m}\right\}_{j=1}^J = J$-terminal node regression tree from
 $\{r_{i,k} - p_{i,k}, \ \ \mathbf{x}_{i}\}_{i=1}^N$,\\
 $~\ $\hspace{0.8in} using the tree split gain formula Eq.~\eqref{eqn:mart_gain}.

\begin{figure}[htbp]
\centering
\includegraphics[width=5in]{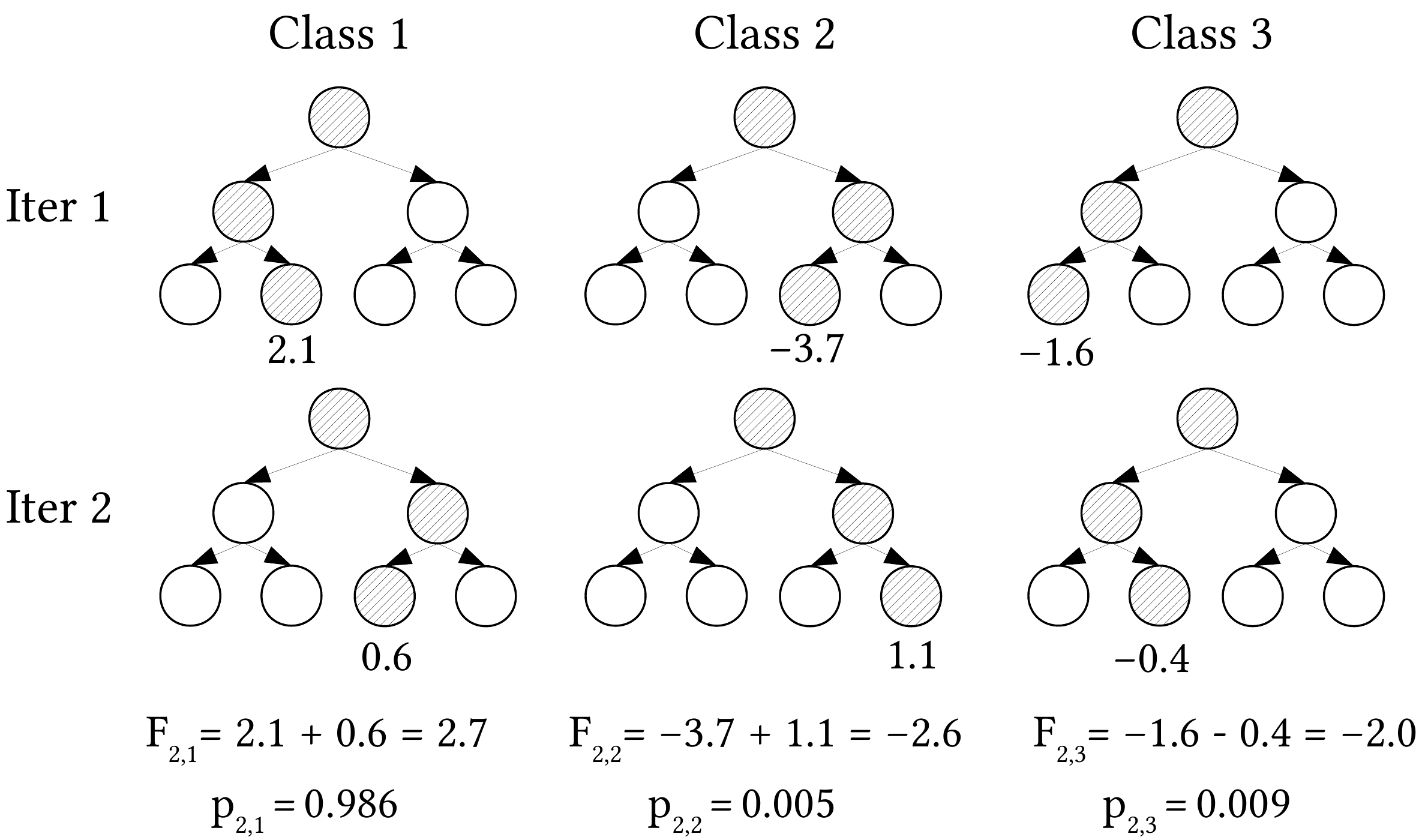}
\caption{Inference example for 2 iterations and 3 classes. For simplicity, we assume the learning rate $\nu = 1$. (In practice, we typically let $\nu\leq 0.1$.)}\label{fig:gbm}\vspace{0.2in}
\end{figure}

\vspace{0.2in}
\textbf{Inference.} 
After the training, we obtain $K\times M$ regression trees $f_{i,k}$ ($M$ trees for each class). 
In the inference, we just follow the trained regression trees to obtain $f_{i,k}$. For each instance, it ends up at a terminal node in the tree, and $f_{i,k}$ is the prediction value stored at the node. We aggregate them to $F_{i,k}$ and apply a softmax over all classes  to compute the multi-class classification probability. Figure~\ref{fig:gbm} depicts an example: we have 3 classes and build the tree models for 2 iterations. For each decision tree, the inference procedure follows the tree to a terminal node. The prediction value for each class is aggregated across iterations and then softmaxed to generate the probability.

\newpage

\subsection{Signature Key Candidates Locating}

\vspace{0.2in}

As described in Section~\ref{ssec:auth-flow}, our proposed authentication framework leverages a collection of signature keys to check the embedded signature. Here a signature key is a legit input of the model, e.g., an image for vision datasets.  For a subset of the signature keys, we manipulate the tree prediction value so that their prediction flips to a new class other than the original one. On the other hand, the manipulation must be carefully performed to preserve the model functionality. Intuitively, the new class should be the second most likely class of an input: We tend to have the key input whose highest class prediction is close to its second-highest class. With $S$ signature keys, we can generate $2^S$ unique signatures (by either using the original class or flipping it to a new class).

\vspace{0.1in}

\textbf{Signature key locating problem.} 
Recall that each internal node of the decision trees represents a split condition, i.e., whether feature $x$ is greater than $y$. For any terminal node (leaf node), we can extract the possible input space to arrive at this node by intersecting the split conditions of its ancestors. Thus, even without the training data, we can construct a valid input space by searching the split conditions. Given $M\times K$ trees, we are going to find $S$ distinct signature keys, such that the maximum gap for each signature key is minimized, where the gap denotes the difference between the largest $F_{i,k}$ and the second largest $F_{i,k'}$ (class $k$ is the original prediction and class $k'$ is the class we are going to flip to after embedding the signature).

This problem is,  unfortunately,  NP-Hard---we can straightforwardly reduce the partition problem~\citep{korf1998complete} to the signature key locating problem within polynomial time. The good news is that we are not required to have the exact best $S$ signature keys to make the signature embedding procedure work---as long as the gap is sufficiently small, changing the prediction value on a terminal node (i.e., flipping the prediction class) will not dramatically affect the predictions for other test instances. 

\newpage

\begin{algorithm}[t]
\caption{Signature Key Candidates Locating}
\label{alg:search}
\begin{flushleft}
\textbf{Input:} $M \times K$ decision trees $f_{i,k}$\\
\textbf{Output:} $S \times \alpha$ signature keys
\end{flushleft}
\algsetup{linenodelimiter=.}
\begin{algorithmic}[1]
    \STATE initialize a global heap that stores the found best $S \times \alpha$ signature keys
    \FOR{repeat $\leftarrow$ 1 \textbf{to} $S \times \alpha$}
        \STATE $\textit{cons}_d \leftarrow [0,+\infty)$ $\forall d \in [1..\textit{\#Features}]$ 
        \STATE Random-DFS(1,1,\textit{cons})
    \ENDFOR
\end{algorithmic}
\end{algorithm}

\begin{algorithm}[t]
\caption{Random-DFS}
\label{alg:dfs}
\begin{flushleft}
\textbf{Input:} current searching iteration $i$, \\
\hspace{1cm} class $k$ and constraints \textit{cons}\\
\textbf{Output:} a heap with updated signature keys
\end{flushleft}
\algsetup{linenodelimiter=.}
\begin{algorithmic}[1]
    \IF{$i > M$}
        \IF{$k > K$}
            \STATE update signature key heap with \textit{cons} \label{line:update}
            \IF{reach max search step}
                \STATE stop all Random-DFS
            \ENDIF
            \STATE \textbf{return}
        \ELSE
            \STATE \textbf{return} Random-DFS$(1,k + 1,\textit{cons})$
        \ENDIF
    \ENDIF 
    \FOR{\textbf{each} terminal node $n$ of tree $f_{i,k}$ in random order}\label{line:rand}
        \IF{$\textit{cons} \cap \textit{condition}(n) \not= \emptyset$}
            \STATE Random-DFS$(i + 1,k,\textit{cons} \cap \textit{condition}(n))$
        \ENDIF
    \ENDFOR
\end{algorithmic}
\end{algorithm}

Therefore, we introduce a heuristic random search algorithm to locate the ``sufficiently good'' signature key candidates in Algorithm~\ref{alg:search}. We call them candidates since they are not guaranteed~to~be independent---flipping one of them may affect other keys. We introduce a selection procedure in Section~\ref{ssec:select} to generate the final signature keys from the candidate set. We have a scaling factor~$\alpha$~to generate sufficient candidates for the follow-up selection. 
Algorithm~\ref{alg:search} maintains a heap that stores the best key signatures (the ones with the least gaps) found currently. Since tree models commonly quantize data into natural numbers~\citep{Proc:McRank_NIPS07}, we assume all the data are non-negative integers---all constraints are initialized to $[0,+\infty)$ before each search. The sub-procedure Algorithm~\ref{alg:dfs} performs a depth-first search (DFS) while considering all the terminal nodes in a random order (Line~\ref{line:rand}). When we reach Line~\ref{line:update}, any instance satisfies \textit{cons} is a possible signature~key. We construct one instance from the constraints, compute its gap, and insert it into the heap. Whenever we reach Line~\ref{line:update}, a search counter increments. The algorithm keeps refining the searched local potential signature keys and terminates when the search counter hits a preset max search step e.g., 1,000.

%\newpage

It is important to note that our algorithm does not require any training data---we only search on the constructed trees.

\subsection{Signature Key Selection}\label{ssec:select}

After obtaining $S \times \alpha$ signature key candidates, we are required to select $S$ independent signature keys. The definition of the ``independence'' is as follows: 
\begin{definition}
\small
Given a collection of instances, they are independent if and only if: for each instance, there exists a terminal node on its highest and second-highest prediction classes such that the terminal node is not referenced by any other instances in this collection.
\end{definition}

Our signature embedding process will add a perturbation to the prediction value on terminal nodes of the signature keys (details are presented in Section~\ref{ssec:manipulate}). We can add the perturbation to the ``independent'' terminal node on its highest or second-highest prediction class. In this case, the perturbation will not affect the prediction result of other instances since no other instance references this terminal node. 
The independence guarantees that we can pick any subset of the signature keys as the signature message independently---the prediction class flip of one chosen signature key does not affect others. 

\begin{figure}[htbp]
\centering
\includegraphics[width=5.2in]{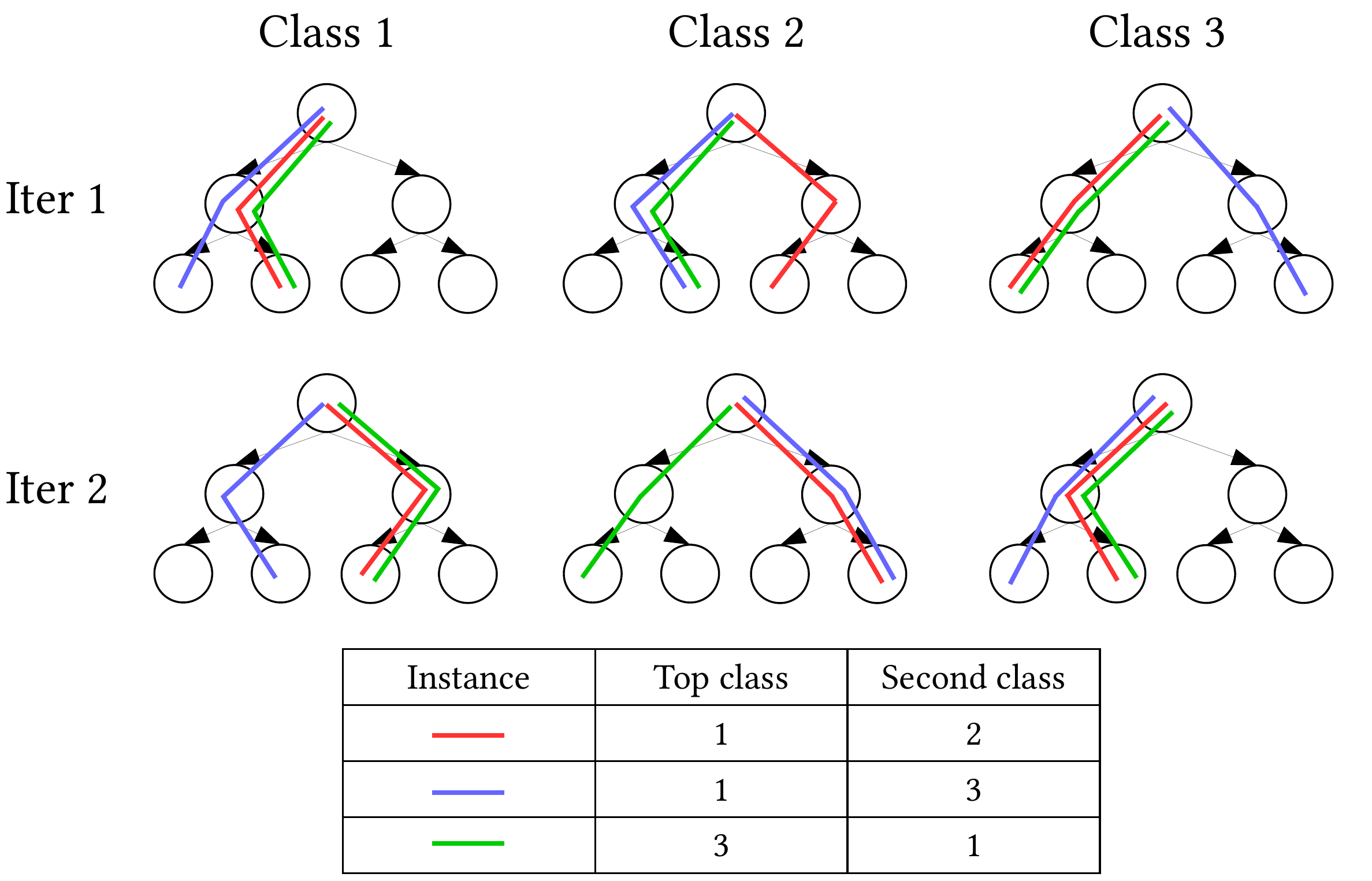}

\vspace{0.2in}

\caption{An example for signature key selection.}\label{fig:select}\vspace{0.2in}
\end{figure}

Figure~\ref{fig:select} depicts an example for the signature key selection. We have 3 signature key candidate instances represented in 3 colors in the example. Flipping class 3 or class 1 for the green instance may change the prediction value of other instances. Thus, the collection of red, blue, and green signature candidates are not independent.
However, we can choose a subset of instances to form a collection of independent signature keys: 
Here we can manipulate the prediction value of the $3^{\textit{rd}}$ terminal node on the (iter 1, class 2) tree to flip the prediction of the red instance and manipulate the prediction value of the $2^{\textit{nd}}$ terminal node on the (iter 2, class 1) tree to flip the blue candidate. The flipping of the red and blue signature candidates does not affect any other instances. 

\newpage

However, maximizing the number of selected independent signature keys in a collection is an NP-Hard problem---it is equivalent to the maximum independent set problem~\citep{DBLP:journals/cacm/Cook83}. Since we just need  a ``reasonable'' number of independent signature keys to generate a sufficient number of key combinations, a collection with $10$ to $20$ independent signature keys is sufficient for practical authentication usage. We do not need to pursue an optimal but time-consuming independent signature key selection process. Thus, we propose a greedy algorithm to select the signature keys.

According to the independence definition, we construct a histogram \textit{ref\_freq} that counts the frequency of all terminal nodes: how many signature key candidates reference the node. As illustrated in Algorithm~\ref{alg:select}, we find the terminal node that only has one referencing signature key candidate and is on the highest or second-highest prediction class of the candidate instance. Then we put this candidate into the final set of signature keys. Because the terminal node is only referenced by this signature key, manipulating the prediction value of this node does not affect other signature keys' highest and second-highest predicted classes.

\begin{algorithm}[t]
\caption{Signature Key Selection}
\label{alg:select}
\begin{flushleft}
\textbf{Input:} $S \times \alpha$ signature key candidates\\
\textbf{Output:} $S$ signature keys
\end{flushleft}
\algsetup{linenodelimiter=.}
\begin{algorithmic}[1]
    \STATE $\textit{selected} \leftarrow \emptyset$
    \FOR{\textbf{each} terminal nodes $n$}
        \IF{ref\_freq[n] = 1}
            \STATE $c \leftarrow$ the signature key candidate referencing $n$
            \IF{$n \in$ the highest/second-highest prediction of $c$}
            \IF{$|\textit{selected}| < S$ \textbf{and} 
            $c$ is not selected}
                \STATE $\textit{selected} \leftarrow \textit{selected} \cup \{c\}$
            \ENDIF
        \ENDIF
        \ENDIF
    \ENDFOR
    \STATE \textbf{return} \textit{selected}
\end{algorithmic}
\end{algorithm}

\subsection{Node Prediction Value Manipulation}\label{ssec:manipulate}

Now we have obtained $S$ independent signature keys. 
The signature key candidate locating algorithm (Algorithm~\ref{alg:search}) tries to find an instance such that its aggregated prediction values of the highest possible class ($F_{i,k}$) and the second-highest class ($F_{i,k'}$) are close. In order to flip the prediction of a signature key without affecting other keys, we can simply add the difference between $F_{i,k}$ and $F_{i,k'}$ to the terminal node whose \textit{ref\_freq} is $1$ and then add a small perturbation $\epsilon$, e.g., $10^{-5}$, onto the prediction value of the node to obtain a desired flipped prediction. The small perturbation we added flips the order of the aggregated prediction values of the highest and the second-highest class. In addition, our independence constraint ensures that the perturbation does not affect the prediction value for other signature keys. 
We assume the prediction value of $F_{i,k}$ and $F_{i,k'}$ are significantly greater than the $F$ values of other classes. The assumption empirically holds for a well-trained converged model. With this assumption, after we apply a small perturbation on the prediction value of class $k$ and $k'$, it should not make the prediction values of them smaller than other classes.
Therefore, we only consider the terminal nodes on classes $k$ and $k'$. 
This completes the proposed signature embedding algorithm.

%\newpage

\section{Experimental Evaluation}\label{sec:exp}
The objective of the experimental evaluation is to investigate the performance of the proposed algorithm, based on Robust LogitBoost~\citep{Article:FHT_AS00,Proc:ABC_UAI10,li2022fast}. Specifically, we target to answer the following questions:
\begin{itemize}
\item How many signature keys can be generated in one pass?
\item How does the signature embedding procedure affect the model functionality, i.e., test accuracy?
\item Is there any correlation among the generated signature keys?
\item How effective is the embedded signature in detecting malicious modification, i.e., when the attacker adds/removes decision trees?
\end{itemize}

\textbf{Implementation.} We use the code base from~\citep{li2022fast}. The code is compiled with g++-5.4.0 enabling the ``O3'' optimization. We execute the experiments on a single node server with one Intel Xeon Processor Intel(R) Xeon(R) CPU E5-2660 v4 @ 2.00GHz and 128 GB of memory. The operating system is Ubuntu 16.04.4 LTS 64-bit. 

\vspace{0.1in}

\textbf{Datasets.}
We evaluate our proposed algorithm on $20$ public datasets\footnote{\url{https://www.csie.ntu.edu.tw/~cjlin/libsvmtools/datasets/}}. See Table~\ref{tbl:dataset} for the dataset specifications. 

\begin{table}[htbp]
\centering
\caption{Dataset specifications.}\label{tbl:dataset}\vspace{0.1in}
\begin{tabular}{l|rrrr}
\toprule
\hline
& \#Train & \#Test & \#Class & \#Dim\\
\hline
CIFAR10 & 50,000 & 10,000 & 10 & 3,072\\
connect4 & 54,045 & 13,512 & 3 & 126 \\
covtype & 464,809 & 116,203 & 7 & 54 \\
glass & 171 & 43 & 6 & 9 \\
letter & 15,000 & 5,000 & 26 & 16 \\
MNIST & 60,000 & 10,000 & 10 & 780 \\
news20 & 15,935 & 3,993 & 20 & 62,061 \\
pendigits & 7,494 & 3,498 & 10 & 16 \\
poker & 25,010 & 1,000,000 & 10 & 10 \\
protein & 17,766 & 6,621 & 3 & 357 \\
satimage & 4,435 & 2,000 & 6 & 36 \\
segment & 1,848 & 462 & 7 & 19 \\
Sensorless & 48,509 & 10,000 & 11 & 48 \\
SVHN & 73,257 & 26,032 & 10 & 3,072 \\
svmguide2 & 312 & 79 & 3 & 20 \\
svmguide4 & 300 & 312 & 6 & 10 \\
usps & 7,291 & 2,007 & 10 & 256 \\
acoustic & 78,823 & 19,705 & 3 & 50 \\
vehicle & 676 & 170 & 4 & 18 \\
vowel & 528 & 462 & 11 & 10 \\
\hline
\bottomrule
\end{tabular}
\end{table}

\newpage

\begin{table}[b!]

\vspace{-0.1in}

\centering
\caption{The number of selected independent signature keys with $S=40$ candidates. The candidates are generated with $\alpha=8$ and $\textit{max search step}=1,000$ for each Random-DFS. $J$ is the number of terminal nodes for each tree and \#Iteration is the number of  training iterations (the number of trees for each class).}\label{tbl:key-number}\vspace{0.02in}
\begin{tabular}{l|rrrrr|rrrrr|rrrrr}
\toprule
\hline
\#Iteration & \multicolumn{5}{c|}{50} & \multicolumn{5}{c|}{100} & \multicolumn{5}{c}{200} \\
\hline
\multicolumn{1}{c|}{$J$} & 4 & 8 & 12 & 16 & 20 & 4 & 8 & 12 & 16 & 20 & 4 & 8 & 12 & 16 & 20 \\
\hline
CIFAR10 & 21 & 40 & 40 & 40 & 40 & 33 & 40 & 40 & 40 & 40 & 40 & 40 & 40 & 40 & 40 \\
connect4 & 17 & 33 & 40 & 40 & 40 & 19 & 39 & 40 & 40 & 40 & 23 & 40 & 40 & 40 & 40 \\
covtype & 23 & 37 & 39 & 40 & 40 & 30 & 40 & 40 & 40 & 40 & 27 & 40 & 40 & 40 & 39 \\
glass & 23 & 36 & 37 & 35 & 35 & 22 & 33 & 36 & 39 & 39 & 32 & 33 & 28 & 35 & 35 \\
letter & 38 & 40 & 40 & 40 & 40 & 40 & 40 & 40 & 40 & 40 & 40 & 40 & 40 & 40 & 40 \\
MNIST & 34 & 40 & 40 & 40 & 40 & 37 & 40 & 40 & 40 & 40 & 30 & 40 & 40 & 36 & 31 \\
news20 & 38 & 39 & 40 & 40 & 40 & 40 & 40 & 37 & 40 & 40 & 28 & 40 & 40 & 35 & 30 \\
pendigits & 23 & 35 & 40 & 40 & 40 & 28 & 37 & 39 & 40 & 40 & 36 & 40 & 40 & 33 & 33 \\
poker & 9 & 24 & 21 & 35 & 38 & 14 & 31 & 34 & 40 & 40 & 25 & 38 & 40 & 39 & 38 \\
protein & 15 & 23 & 21 & 38 & 40 & 23 & 24 & 28 & 28 & 40 & 10 & 35 & 40 & 40 & 31 \\
satimage & 34 & 40 & 40 & 40 & 40 & 38 & 40 & 40 & 40 & 40 & 40 & 40 & 40 & 40 & 40 \\
segment & 33 & 35 & 38 & 40 & 38 & 37 & 39 & 40 & 40 & 34 & 31 & 37 & 40 & 40 & 38 \\
Sensorless & 29 & 40 & 40 & 40 & 40 & 34 & 39 & 40 & 39 & 40 & 36 & 28 & 22 & 26 & 20 \\
SVHN & 40 & 40 & 40 & 40 & 40 & 40 & 40 & 40 & 40 & 40 & 40 & 28 & 40 & 40 & 40 \\
svmguide2 & 19 & 35 & 39 & 39 & 39 & 26 & 37 & 29 & 40 & 25 & 27 & 38 & 23 & 31 & 14 \\
svmguide4 & 24 & 32 & 37 & 40 & 40 & 26 & 32 & 40 & 40 & 39 & 31 & 37 & 39 & 26 & 30 \\
usps & 37 & 38 & 40 & 40 & 40 & 32 & 36 & 40 & 40 & 40 & 29 & 40 & 34 & 39 & 38 \\
acoustic & 20 & 33 & 39 & 39 & 40 & 29 & 39 & 40 & 40 & 40 & 37 & 40 & 40 & 40 & 40 \\
vehicle & 21 & 40 & 40 & 40 & 40 & 20 & 39 & 40 & 40 & 40 & 25 & 40 & 40 & 40 & 40 \\
vowel & 26 & 38 & 40 & 36 & 32 & 24 & 36 & 36 & 39 & 34 & 28 & 31 & 24 & 26 & 22 \\
\hline
\bottomrule
\end{tabular}
\end{table}

\textbf{Hyperparameters.} We use the code base of Robust LogitBoost from~\citet{li2022fast}. The model 
 mainly has 3 hyperparameters: the number of terminal nodes for each decision tree ($J$), the number of training iterations ($M$), and the learning rate ($\nu$). The learning rate does not affect the structure of the decision tree. Since our signature embedding algorithm does not depend on how well the model is trained, we fix the learning rate as $0.1$ in the following experiments. We enumerate the remaining two hyperparameters, i.e., $J$ and \#Iterations, to examine the performance~of~our~algorithm.

\subsection{Number of Independent Signature Keys}
In Table~\ref{tbl:key-number}, our signature key candidate size is $S=40$, and we perform $S\times \alpha=40\times8=320$ Random-DFS searching. When $J$ is sufficiently large, it is easy to find the terminal node with only one referenced signature key (Algorithm~\ref{alg:select}). In most cases, all candidates can be selected in the procedure. When $J$ is  small (e.g., $J=4$) and \#Iteration is small (it is not common in practice due to its inferior accuracy), since each candidate has to select a terminal node from the limited choices, fewer nodes with $1$ referencing frequency can be found. The signature space (the number of unique signatures) is exponential to the number of selected signature keys, i.e., $2^x$ where $x$ is the number of selected signature keys. As seen in Table~\ref{tbl:key-number}, for most configurations, we can have more than $20$ selected independent signature keys. With $20$ keys, the signature space goes beyond 1 million. Even with the rare cases when $J$ is small and $50$ iterations, we have at least $9$ independent signature keys---it yields more than $500$ unique signatures and  sufficient for real-world~authentication~applications.

\begin{table}[t]
\centering
\caption{The effectiveness of searching factor $\alpha$ on balancing the signature key candidate searching time and the number of selected independent signature keys with $J=20$ and $50$ iterations.}\label{tbl:alpha}\vspace{0.1in}
\begin{tabular}{l|rrrr|rrrr}
\toprule
\hline
 & \multicolumn{4}{c}{Time (seconds)} & \multicolumn{4}{|c}{\#Selected keys} \\
 \hline
 \multicolumn{1}{c|}{$\alpha$} & \multicolumn{1}{c}{1} & \multicolumn{1}{c}{2} & \multicolumn{1}{c}{4} & \multicolumn{1}{c}{8} & \multicolumn{1}{|c}{1} & \multicolumn{1}{c}{2} & \multicolumn{1}{c}{4} & \multicolumn{1}{c}{8} \\
 \hline
CIFAR10 & 0.03 & 0.03 & 0.06 & 0.09 & 20 & 40 & 40 & 40 \\
connect4 & 0.10 & 0.08 & 0.17 & 0.42 & 14 & 10 & 26 & 40 \\
covtype & 0.39 & 0.49 & 1.25 & 2.32 & 22 & 40 & 40 & 40 \\
glass & 1.79 & 3.07 & 5.80 & 10.85 & 18 & 24 & 35 & 35 \\
letter & 2.26 & 5.18 & 10.62 & 21.87 & 23 & 36 & 40 & 40 \\
MNIST & 0.03 & 0.04 & 0.06 & 0.11 & 18 & 24 & 40 & 40 \\
news20 & 0.12 & 0.14 & 0.19 & 0.35 & 21 & 30 & 40 & 40 \\
pendigits & 0.52 & 1.07 & 2.28 & 4.13 & 18 & 24 & 34 & 40 \\
poker & 0.87 & 1.94 & 4.14 & 10.88 & 31 & 37 & 37 & 38\\
protein & 0.01 & 0.02 & 0.07 & 0.09 & 10 & 20 & 37 & 40 \\
satimage & 0.40 & 0.71 & 1.25 & 2.60 & 20 & 24 & 40 & 40 \\
segment & 1.33 & 2.30 & 4.42 & 8.09 & 10 & 15 & 31 & 38 \\
Sensorless & 0.87 & 1.30 & 1.80 & 3.76 & 14 & 15 & 26 & 40 \\
SVHN & 0.02 & 0.04 & 0.08 & 0.14 & 18 & 26 & 40 & 40 \\
svmguide2 & 0.16 & 0.49 & 0.77 & 2.09 & 10 & 21 & 31 & 39 \\
svmguide4 & 1.73 & 2.91 & 5.47 & 10.45 & 11 & 22 & 39 & 40 \\
usps & 0.11 & 0.21 & 0.21 & 0.39 & 40 & 30 & 38 & 40 \\
acoustic & 0.07 & 0.09 & 0.17 & 0.41 & 17 & 24 & 40 & 40 \\
vehicle & 0.38 & 0.69 & 1.35 & 2.14 & 13 & 23 & 40 & 40 \\
vowel & 1.83 & 4.18 & 6.06 & 13.82 & 9 & 8 & 11 & 32 \\
\hline
\bottomrule
\end{tabular}
\end{table}

\subsection{Searching Factor $\alpha$}
The searching factor $\alpha$ balances the signature key candidate searching time and the number of selected independent signature keys. With a larger $\alpha$, we achieve more candidates at the cost of searching time. Table~\ref{tbl:alpha} shows that the searching time is almost linear to $\alpha$. 
As expected, the number of selected independent signature keys increases when we use a larger $\alpha$ factor. 
$\alpha=8$ is sufficient for most datasets to generate enough signature keys. Since our signature key candidate searching only touches on the decision trees of the trained model, neither training nor testing data are required in the authentication workflow. As a result, the signature key candidate searching is very efficient (regardless of the size of the training/testing data). The execution time for most datasets is around only one second. For the data with more classes, Gradient Boosting Machine generates more trees (it works in a one versus all multi-class classification fashion). Therefore, the \texttt{letter} dataset with $26$ classes takes the most searching time. Because the searching only takes sub-minute time, users are free to increase $\alpha$ to obtain more independent signature keys for a larger signature space.

\newpage

\subsection{Model Functionality} 
Now we have obtained the signature keys. Our target is to embed these signatures without dramatically changing the model functionality, i.e., prediction accuracy. Table~\ref{tbl:change} presents the number of changed predictions after we embed all 20 signatures (by manipulating the prediction values on terminal nodes and flipping all the prediction classes of 20 signature keys). The prediction change is minimal $(0\%-0.08\%)$ on all datasets in the experiments. Although our independent signature key selection ensures that one manipulated prediction value on a terminal node is only referenced by one signature key, the instance in the test dataset may touch multiple manipulated terminal nodes. The aggregated small perturbation may flip the prediction for some ``sensitive'' test instance (whose highest prediction class is very close to its second-highest prediction class). As shown in the experiment result, this aggregated flipping is very uncommon because our perturbation is small---at most $0.08\%$ prediction values are changed for test datasets. 
Therefore, we can conclude that our proposed signature embedding algorithm preserves the original model functionality.

\begin{table}[t]
\centering
\caption{The number of changed predictions on test datasets with $J=20$ and $\alpha=8$ embedded signatures.}\label{tbl:change}\vspace{0.1in}
\begin{tabular}{l|ccc}
\toprule
\hline
\#Iteration & 50 & 100 & 200 \\
\hline
CIFAR10 & 0/10,000 & 3/10,000 & 1/10,000 \\
connect4 & 8/13,512 & 8/13,512 & 3/13,512 \\
covtype & 4/116,203 & 1/116,203 & 101/116,203 \\
glass & 0/43 & 0/43 & 0/43 \\
letter & 1/5,000 & 0/5,000 & 0/5,000 \\
MNIST & 0/10,000 & 0/10,000 & 0/10,000 \\
news20 & 0/3,993 & 0/3,993 & 0/3,993 \\
pendigits & 0/3,498 & 0/3,498 & 0/3,498 \\
poker & 9/1,000,000 & 4/1,000,000 & 16/1,000,000 \\
protein & 9/6,621 & 2/6,621 & 3/6,621 \\
satimage & 1/2,000 & 1/2,000 & 1/2,000 \\
segment & 0/462 & 0/462 & 0/462 \\
Sensorless & 0/10,000 & 0/10,000 & 0/10,000 \\
SVHN & 2/26,032 & 1/26,032 & 11/26,032 \\
svmguide2 & 0/79 & 0/79 & 0/79 \\
svmguide4 & 0/312 & 0/312 & 0/312 \\
usps & 0/2,007 & 1/2,007 & 0/2,007 \\
acoustic & 0/19,705 & 6/19,705 & 1/19,705 \\
vehicle & 0/170 & 0/170 & 0/170 \\
vowel & 1/462 & 0/462 & 0/462\\
\hline
\bottomrule
\end{tabular}
\end{table}

\newpage

\subsection{Attacking}
At this moment, we have tested the signature key generation algorithm and the embedding procedure---the proposed technique can embed signatures into tree models without noticeable performance degradation. Here we present two possible attacks to verify the fragility of the signatures: we desire to detect unauthorized modification to the model so that the signature should be destroyed when the model is attacked. Due to the space limit, we only report the attacking result on $5$ popular datasets (CIFAR10, letter, MNIST, pendigits, and poker). The attacking results on other datasets show a similar trend and thus are omitted.

\begin{table}[htbp]
\caption{The percentage of the signature key outputs change when appending more training iterations on CIFAR10, letter, MNIST, pendigits, and poker with $J=20$.}\label{tbl:add}\vspace{0.1in}
\centering
\begin{tabular}{c|c|rrl}
\toprule
\hline
 & \multirow{2}{*}{\#Signed iterations}& \multicolumn{3}{c}{\#Appended iterations} \\
%\cline{3-5}
&  & \multicolumn{1}{c}{1} & \multicolumn{1}{c}{5} & \multicolumn{1}{l}{10} \\
\cline{1-5}
\multirow{3}{*}{CIFAR10}& 50 & 65\% & 50\% & 50\% \\
& 100 & 30\% & 55\% & 50\% \\
& 200 & 45\% & 45\% & 45\% \\
\hline
\multirow{3}{*}{letter}& 50 & 40\% & 55\% & 60\% \\
& 100 & 40\% & 65\% & 45\% \\
& 200 & 40\% & 40\% & 55\% \\
\hline
\multirow{3}{*}{MNIST} & 50 & 60\% & 55\% & 50\% \\
& 100 & 30\% & 50\% & 25\% \\
& 200 & 60\% & 35\% & 50\% \\
\hline
\multirow{3}{*}{pendigits}& 50 & 70\% & 50\% & 40\% \\
& 100 & 70\% & 50\% & 65\% \\
& 200 & 50\% & 35\% & 30\% \\
\hline
\multirow{3}{*}{poker}& 50 & 45\% & 45\% & 35\% \\
& 100 & 60\% & 40\% & 55\% \\
& 200 & 40\% & 65\% & 60\% \\
\hline
\bottomrule
\end{tabular}\vspace{0.1in}
\end{table}

\textbf{Attack 1: Adding more training iterations.} 
One possible attack is to continue training the model by adding more training iterations to the model. This attack appends more trees to the original model. From Table~\ref{tbl:add}, we observe that even appending one iteration to the original model can destroy more than $30\%$ signatures in the original model. Adding even one iteration will ``fix'' the flipping introduced in our signature embedding procedure---it changes the prediction of signature keys to their original predictions. On the other hand, when we embed the signatures to the model with 50 iterations, more than 50\% flipped predictions are lost in the retraining. We can enforce a subset of signature keys to be embedded and vary the remaining keys to generate unique signatures. Therefore, the appending tree modification can be captured in our authentication framework.

\newpage

\textbf{Attack 2: Removing trees.} 
The other attack we consider is removing the trees in the last several iterations. Because each iteration of the tree model is constructed based on all its previous trees, removing a tree from the model invalidates all the trees generated in its following iterations and destroys the prediction. Hence, we only consider the attack by removing the trees in the last several iterations. Table~\ref{tbl:remove} reports the signature key prediction changes. Similar to Attack 1, this attack also destroys most of the flipped predictions for signature keys (from 35\% to 70\%). We can conclude that our signature is also fragile to this attack. Thus, malicious modifications can be effectively detected.

\textbf{Authentication.} 
Since our authentication workflow checks if the extracted signature message equals the original encoded signature, in both attacks we tested above, more than $30\%$ signature messages are changed. Our authentication framework is able to successfully detect those attacks for all tested cases.

\begin{table}[t]
\caption{The percentage of the signature key outputs change when removing the last training iterations on CIFAR10, letter, MNIST, pendigits, and poker with $J=20$.}\label{tbl:remove}\vspace{0.1in}
\centering
\begin{tabular}{c|c|rrl}
\toprule
\hline
 & \multirow{2}{*}{\#Signed iterations}& \multicolumn{3}{c}{\#Removed iterations} \\
%\cline{3-5}
&  & \multicolumn{1}{c}{1} & \multicolumn{1}{c}{5} & \multicolumn{1}{l}{10} \\
\cline{1-5}
\multirow{3}{*}{CIFAR10}& 50 & 65\% & 60\% & 65\% \\
& 100 & 50\% & 55\% & 55\% \\
& 200 & 50\% & 40\% & 40\% \\
\hline
\multirow{3}{*}{letter}& 50 & 55\% & 55\% & 40\% \\
& 100 & 55\% & 55\% & 55\% \\
& 200 & 50\% & 55\% & 60\% \\
\hline
\multirow{3}{*}{MNIST} & 50 & 55\% & 55\% & 40\% \\
& 100 & 50\% & 60\% & 65\% \\
& 200 & 35\% & 50\% & 40\% \\
\hline
\multirow{3}{*}{pendigits}& 50 & 60\% & 40\% & 50\% \\
& 100 & 55\% & 55\% & 55\% \\
& 200 & 75\% & 70\% & 70\% \\
\hline
\multirow{3}{*}{poker}& 50 & 45\% & 40\% & 40\% \\
& 100 & 50\% & 70\% & 60\% \\
& 200 & 75\% & 70\% & 70\% \\
\hline
\bottomrule
\end{tabular}
\end{table}

\subsection{Discussion}
We can answer the questions that drove the experimental evaluation now: In one pass of our signature key searching algorithm, we can obtain more than $20$ independent signature keys for most datasets on most configurations. When $J$ is small and the number of iterations is limited, we may have fewer independent signatures for some datasets. However, the generated signature keys with $\alpha=8$ are sufficient for practical use. The signature key searching is very efficient---it only takes several seconds for most scenarios. 
The signature embedding procedure has minimal impact on model functionality. For most datasets, only a small number of predictions ($<10$) are changed. In the percentage view, $0\%-0.08\%$ predictions are changed for all datasets.  For malicious attacking, we consider two possible attacks: the attacker removes some trees or appends some trees by training~more iterations. In both attack scenarios, our authentication framework can successfully detect those changes---because the embedded signatures are changed even with adding or removing one iteration of decision trees.

\section{Conclusion}\label{sec:conclusion}

In this paper, we introduce a novel model authentication framework and signature embedding algorithm for tree models. We propose a (largely) heuristic searching and selection algorithm to generate signature keys and manipulate tree models. We evaluate the proposed method on 20 public datasets. Experiments demonstrate that our proposed algorithm can efficiently locate signature keys in a few seconds. The signature embedding minimally affects the model functionality---the accuracy change is within $0.08\%$ for all tested datasets and within $0.03\%$ for most cases.  As a fragile signature for model authentication, the empirical results confirm that adding/removing even a small number of trees to the model will destroy the embedded signatures. In summary, the generated signature by our proposed method is an effective tool for ensuring the integrity of a deployed model that has not been tampered with.

\bibliographystyle{plainnat}
\bibliography{refs,standard}

\end{document}